\documentclass[12pt]{article}

\usepackage{amssymb,amsmath}
\usepackage{cite}

\usepackage{graphicx}

\newcommand{\be}{\begin{equation}}
\newcommand{\ee}{\end{equation}}

\newcommand{\bt}{\beta}

\newcommand{\ra}{\rightarrow}

\newcommand{\gm}{\gamma}

\newcommand{\lbd}{\lambda}

\newcommand{\cH}{{\cal H}}

\newcommand{\cL}{{\cal L}}

\newcommand{\lgl}{\langle}
\newcommand{\rgl}{\rangle}

\begin{document}

\begin{center}

{\Large{\bf How brains make decisions} \\ [5mm]

V.I. Yukalov$^{1,2}$ and D. Sornette$^{1,3}$} \\ [3mm]

{\it
$^1$Department of Management, Technology and Economics, \\
ETH Z\"urich, Swiss Federal Institute of Technology, \\
Z\"urich CH-8092, Switzerland \\ [3mm]

$^2$Bogolubov Laboratory of Theoretical Physics, \\
Joint Institute for Nuclear Research, Dubna 141980, Russia \\ [3mm]

$^3$Swiss Finance Institute, c/o University of Geneva, \\
40 blvd. Du Pont d'Arve, CH 1211 Geneva 4, Switzerland}

\end{center}

\vskip 3cm

\begin{abstract}

This chapter, dedicated to the memory of Mino Freund, summarizes
the Quantum Decision Theory (QDT) that we have developed in a series of publications 
since 2008. We formulate a general mathematical scheme of how decisions are taken, 
using the point of view of psychological and cognitive sciences, without touching 
physiological aspects. The basic principles of how intelligence acts are discussed. 
The human brain processes involved in decisions are argued to be principally 
different from straightforward computer operations. The difference lies in the 
conscious-subconscious duality of the decision making process and the role of emotions 
that compete with utility optimization. The most general approach for characterizing 
the process of decision making, taking into account the conscious-subconscious duality, 
uses the framework of functional analysis in Hilbert spaces, similarly to that used 
in the quantum theory of measurements. This does not imply that the brain is a 
quantum system, but just allows for the simplest and most general extension of
classical decision theory.  The resulting theory of quantum decision making, based 
on the rules of quantum measurements, solves all paradoxes of classical decision making, 
allowing for quantitative predictions that are in excellent agreement with experiments. 
Finally, we provide a novel application by comparing the predictions of QDT with 
experiments on the prisoner dilemma game. The developed  theory can serve as a guide 
for creating artificial intelligence acting by quantum rules.    
 
\end{abstract}

\newpage

\section{What is brain intelligence}

The brain is the center of the nervous system in all vertebrates and most 
invertebrates. Only a few invertebrates, such as sponges, jellyfish, sea squirts,
and starfish do not have one, though they have diffuse neural tissue. The brain 
of a vertebrate is the most complex organ of its body. In a typical human, the 
cerebral cortex is estimated to contain $86 \pm 8$ billion neurons 
\cite{Herculano2009}, each connected by synapses to several thousand other neurons. 

The functioning of the brain can be considered from two different perspectives, 
physiological and psychological. We do not touch here the physiological side of 
the problem that is studied in neurobiology, medicine, and is also modeled by
neuron networks \cite{Shepherd_1,Kandel_2,Sporns_3}. Our aim is to model
the functioning of the psychological brain, which is studied in cognitive sciences.     

The ability of the brain to take decisions is termed intelligence. There exist 
numerous and rather lengthy discussions attempting to describe what intelligence is 
\cite{Luria_4,Elkind_5,Sternberg_6,Das_7,Wake_8,Rishardson_9,Sternberg_10,Stanovich_11}. 
Summarizing these discussions, the basic feature of intelligence, which can be 
accepted as its brief definition, is {\it the ability of adaptation to the environment 
by the process of making optimal decisions}. This implies that the notion of 
intelligence is foremost the ability of making decisions. It is generally accepted 
that humans possess the highest level of intelligence in the animal kingdom. But 
animals also are able to take decisions, to adapt to their environment and to solve 
problems \cite{Coren_12}. Thence, animals also possess intelligence. This concerns all 
animals, such as birds, fish, reptiles, amphibians, and insects. Moreover, other 
living beings, say plants, in some sense, do adapt to surrounding by making 
decisions \cite{Trewavas_13}. Therefore, we need to accept that, to some degree, 
all alive beings have a kind of intelligence, since all of them adjust to 
their environment by reacting to external signals. Thus, one can talk of the intelligence 
of plants, fungi, bacteria, protista, amoebae, algae, and so on. In that sense, 
any entity that is able to take decisions, adapting to surrounding signals, can be 
assumed to have something like intelligence. If such an entity that is able to take 
optimal decisions is created by humans, it is called artificial intelligence
\cite{Canny_14}. 

In the following, we shall be mostly concerned with the functioning of the human 
brain, though many parts of our considerations could be applied to the functioning
of the brains and nervous systems of other alive beings. The human brain, being the 
most developed and complex, exhibits in the most explicit way the features that 
could be met in the behavior of other animals. The aim of this paper is to 
demonstrate that the human brain makes decisions in a rather intricate way that
cannot be described by the classical utility or prospect theory used in economics. 
We argue that decisions made by brains are not the same as straightforward 
computer-like calculations. Human decisions are based on the functioning of and 
interplay between conscious as well as subconscious processes of the brain. This 
complex behavior can be represented by the techniques of quantum theory,
which seems to be the most general and simplest framework for realistically 
characterizing the decision making process of human brains.   

The plan of the paper is as follows. In Sec. 2, we recall how decisions are 
supposed to be made by fully rational decision makers who evaluate the utility 
of prospects and choose the one with the largest utility. Such a strictly deterministic 
behavior is a strong simplification of the reality. Empirical observations show 
that there always exists a distribution of choices made by different subjects,
rather a single optimal behavior. Even the same subject, under varying conditions 
or time, can make different choices when confronted with the same set of competing 
prospects. 

This implies that the first step towards a realistic representation  
of decision making is the reformulation of classical utility theory within a 
probabilistic framework, which is accomplished in Sec. 3. Analyzing the signals, the 
subject formulates a set of possible actions, $\pi_j$, termed prospects that are 
weighted with probabilities $p(\pi_j)$. Taking a decision means the selection of an 
optimal prospect $\pi_*$ characterized by the largest probability, though other 
prospects can also be chosen, with lower probabilities, that is, with lower frequency. 
The possible actions are always weighted with a probability distribution. This 
describes the probability weighted diversity of choices among a population of similar 
decision makers. There always exists a probability that some of the members choose 
one prospect, while others choose another prospect, although the majority prefers the 
optimal prospect. This is the essence of the probability weight that is associated 
with the frequentist interpretation, which defines the fraction of those who choose 
the related prospect.

Although the probabilistic utility theory that we introduce in Sec. 3 generalizes 
the standard deterministic utility theory, it does not take into account that real 
decision makers are not fully rational. Moreover, they experience a variety of emotions 
and behavioral biases. As a result, decisions are taken not by a simple evaluation 
of utilities but are essentially influenced by these biases and emotions. In taking 
decisions, two brain processes are involved, conscious and subconscious. This dual 
functioning of the brain makes its principally distinct from the straightforward 
calculations by a computer, as is discussed in Sec. 4.

To take into account this complex dual behavior, Sec. 5 presents a generalization 
of decision theory, which invokes the techniques of the quantum theory of measurements. 
The developed Quantum Decision Theory (QDT) contains none of the paradoxes that are 
so numerous in classical decision making. Importantly, we show that classical decision 
theory constitutes a particular case of QDT. The latter reduces to the former under 
a process that can be called ``decoherence'', which describes how the addition of 
reliable information decreases the emotional component of a decision, thus making 
it more and more controlled by the rational utility component.

To illustrate how QDT describes how decisions are made, avoiding the paradoxes of 
classical decision making and providing quantitative predictions, we treat in Sec. 6 
the prisoner dilemma game.

Section 7 summarizes the results, stressing that the developed QDT is, to our knowledge, 
the sole decision theory that not merely removes classical paradoxes, but provides 
{\it quantitative} predictions, with {\it no adjustable parameters}, which are in good 
agreement with empirical observations.       

Concluding this introduction, our main hypothesis is that the brain makes decisions 
through a procedure that is similar to quantum measurements. This does not require 
the brain to be a quantum object, but merely takes into account the dual nature of 
the decision process, involving both conscious logical evaluations as well as  
subconscious intuitive feelings. This chapter summarizes the Quantum Decision Theory 
(QDT) that we have developed in a series of publications 
\cite{Yukalov_44,Yukalov_45,Yukalov_46,Yukalov_47,Yukalov_48,Yukalov_49}.
We also provide a novel application  on the prisoner dilemma game, comparing the 
predictions of QDT with experiments.

\section{Choosing a prospect on fully rational grounds}

Assuming that the subject is fully rational and possesses the whole necessary 
information for making decisions, it is reasonable to suppose that such decisions are 
based on the evaluation of the utility of the results following the 
corresponding action. This is the central assumption of expected utility theory,
which prescribes a normative framework on how decisions are made. The basic mathematical 
rules of expected utility theory have been compiled by von Neumann and Morgenstern
\cite{Neumann_15} and Savage \cite{Savage_16}. Below, we give a brief sketch of the 
main features of utility theory in order to introduce the terminology to be used 
in the following sections, where the generalizations of this theory will be considered.

The outcomes of actions, that is, the consequences of events, are measured by payoffs 
composing a set 
\be
\label{1}
 \mathbb{X} \equiv \{ x_n \in \mathbb{R} : \; n = 1,2,\ldots,N_{out} \} \; .
\ee
The number of outcomes $N_{out}$ can be as small as two or asymptotically large. 
Positive outcomes correspond to gains, while negative ones to losses. Payoffs $x_n$ can
come with different probabilities $p_j(x_n)$, being labeled by an index 
$j = 1,2,\ldots, L$, and satisfying the normalization condition
\be
\label{2}
 \sum_{n=1}^{N_{out}} p_j(x_n) = 1 \; , \qquad 0 \leq p_j(x_n) \leq 1 \;  .
\ee
The ensemble of payoffs and their probabilities is called a lottery, or a prospect
\be
\label{3}
 \pi_j = \{ x_n , p_j(x_n) : \; n= 1,2,\ldots, N_{out} \} \;  .
\ee
One also uses the notion of compound lotteries that are the linear combinations of 
a given set of lotteries, with the same payoffs and with the linear combinations 
of the related weights. 

There can exist several prospects forming a family
\be
\label{4}
\cL = \{ \pi_j : \; j= 1,2,\ldots, L \} \;   .
\ee
The task of decision making is to decide between the prospects $\pi_j$, choosing one 
out of the given family.     

The choice involves the classification of outcomes according to their utility for the
decision maker. One defines a utility function $u(x): \mathbb{X} \ra \mathbb{R}$ that
can also be called pleasure function, satisfaction function, or profit function. By
definition, the utility function is nondecreasing (more is always preferred), so that 
$u(x_1) \geq u(x_2)$ for $x_1 \geq x_2$ and concave (diminishing marginal utility), 
such that $u(\alpha_1 x_1 + \alpha_2 x_2) \geq \alpha_1 u(x_1) + \alpha_2 u(x_2)$ for 
non-negative $\alpha_i$'s normalized to one. The first derivative $u'(x) \equiv du(x)/dx$ 
is termed the marginal utility that is non-negative for a non-decreasing function. 
The second derivative $u''(x) \equiv d^2u(x)/dx^2$ is non-positive for a concave 
function. Hence, the marginal utility $u'(x)$ does not increase. This implies that, 
with increasing payoff $x$, the utility function decelerates. Such a function is termed 
risk averse \cite{Pratt_17,Arrow_18}, since a sure payoff is always preferred
to different random payoffs with the same mean value. The risk aversion can be 
captured by the so-called degree of risk aversion $r(x) \equiv - u''(x)/u'(x)$, which 
is non-negative. Examples of utility functions are linear, power-law, logarithmic or 
exponential functions. Usually, the utility of nothing is set to zero, $u(0) = 0$, but 
the absolute utility level is inconsequential.

Generally, a payoff $x_n$ can be either positive, representing a gain, or negative
corresponding to a loss. Strictly speaking, it is impossible to lose something, while 
having nothing. Even the poorest person can lose a gamble and go in debt, having an 
instantaneous negative net worth. However, taking into account the value of future incomes 
gives in general a positive net value and the debt then constitutes a loss of a part of 
future incomes. There can be however situations where debt reaches levels beyond the 
most optimistic expectations of future incomes, so that one has lost what one did not 
own now or will ever have in the future. In its encyclopedic review of the history of 
debt in human societies, Graeber documents that such situations were quite common  
\cite{Graeber}. They were usually followed  by slavery (and are still in various
explicit or disguised forms followed by some kind of slavery), where the person in debt
sells his children, wife or himself.  A loss is then backed up by the ultimate reservoir 
of wealth, being stored in the value of oneself \cite{Graeber}. Formally, this implies 
that, before losing $x_n$, one has an initial given amount $x_0 \geq x_n$. Then, 
shifting all payoffs by $x_0$, one can redefine the lottery so that all its payoffs 
be non-negative.    

Each prospect is characterized by the expected utility
\be
\label{5}
 U(\pi_j) = \sum_{n=1}^{N_{out}} u(x_n) p_j(x_n)  .
\ee
This notion was introduced by Bernoulli \cite{Bernoulli_19} and an axiomatic theory 
was developed by von Neumann and Morgenstern \cite{Neumann_15}, where the payoff 
weights were treated as objective. Savage \cite{Savage_16} extended the notion to 
subjective probabilities evaluated by decision makers. 

Expected utility can be interpreted either as a cardinal or ordinal quantity. 
Cardinal utility is assumed to be precisely measured and the magnitude of the 
measurement is meaningful. It can be measured in some chosen units, similarly to how
distance is measured in meters, or time in hours, or weight in kilograms. However,
such a definition in precise units is often impossible and, actually, not necessary.
It is sufficient to interpret the expected utility as ordinal utility, for which
its precise magnitude is not important, but the magnitude of the ratios between 
different utilities carries sufficient meaning.

The prospect uncertainty is described by the prospect variance
\be
\label{6}
{\rm var}(\pi_j) \equiv \frac{1}{N_{out}} \; \sum_{n=1}^{N_{out}} \left \{
x_n^2 p_j(x_n) - [\overline x(\pi_j) ]^2 \right \} \;  ,
\ee
whose square root can also be called the prospect volatility or spread. We
have used the following notation for the prospect mean
\be
\label{7}
\overline x(\pi_j) \equiv \frac{1}{N_{out} } \sum_{n=1}^{N_{out}} x_n p_j(x_n)~.
\ee  

The ordering of the prospects is prescribed by the relations between their 
expected utilities. One says that a prospect is preferable to another one if its
utility is larger than that of the latter. Two prospects are termed indifferent
when their utilities coincide. The properties of the utility function $u(x)$
prescribe the properties of the expected utility.

\vskip 2mm 
(i) {\it Completeness}: For any two prospects $\pi_1$ and $\pi_2$, 
one of the following relations necessarily holds, either $\pi_1 = \pi_2$, or 
$\pi_1 < \pi_2$, or $\pi_1 > \pi_2$, or $\pi_1 \leq \pi_2$, or $\pi_1 \geq \pi_2$, 
understood as the corresponding relations between their expected utilities. 

(ii) {\it Transitivity}: For any three prospects, such that $\pi_1 \leq \pi_2$ and
$\pi_2 \leq \pi_3$, it follows that $\pi_1 \leq \pi_3$.  

(iii) {\it Continuity}: For any three prospects ordered so that 
$\pi_1 \leq \pi_2 \leq \pi_3$, there exists $\alpha \in [0,1]$ such that 
$\pi_2 = \alpha \pi_1 + (1 - \alpha) \pi_3$. 

(iv) {\it Independence}: For any $\pi_1 \leq \pi_2$ and an arbitrary $\pi_3$, there 
exists $\alpha \in [0,1]$ such that 
$\alpha \pi_1 + (1 - \alpha) \pi_3 \leq \alpha \pi_2 + (1 - \alpha) \pi_3$.

\vskip 2mm
The central aim of expected utility theory is to calculate the expected 
utilities for all prospects from the given family $\mathcal{L}$ and, comparing their
values, to find the prospect possessing the largest utility. Then the decision is 
taken by selecting this prospect corresponding to the largest utility, which is 
called the {\it most useful prospect}. The decision making scheme based
on expected utility theory is given in Fig. 1.

\section{Probabilistic approach to expected utility theory}

According to the expected utility theory delineated above, the choice of a 
prospect is with certainty prescribed by the utility of the prospects. This
theory is deterministic, since the choice, with probability one, is required
to correspond to the prospect with the largest expected utility. Such a 
completely deterministic formulation contradicts the known empirical facts demonstrating 
that, under the same conditions, different persons often choose different 
prospects. Of course, one could salvage the deterministic theory by
introducing heterogenous utility functions that describe the variety of tastes
of different people \cite{McFadden1981}. While this captures the evident observation 
that tastes exhibit some heterogeneity, extending utility theory to heterogeneous
or random utility theory comes at the cost of a proliferation of parameters, 
making the approach descriptive at best, while being non-parsimonious and non predictive.
An even more convincing attack to the deterministic approach comes from the 
observation that the same person, under the same conditions, may choose 
different prospects at different times. This ``intra-observer variation'' has
been largely documented in the medical literature \cite{Rutkow,Nielsenetal90}.
This suggests to view the brain of a decision maker as deliberating on the set 
of admissible prospects and evaluating them by involving some probabilistic weighting. 
This is the motivation to reformulate utility theory by generalizing it to a 
probabilistic approach.       

The probabilistic weighting of prospects can be formalized by invoking the 
principle of minimal information that allows one to find a probability distribution
under the minimal given information. The idea of this principle goes back to 
Gibbs \cite{Gibbs_20,Gibbs_21,Gibbs_22}, who formulated it as a conditional 
maximization of entropy under the given set of constraints. This principle is 
widely used in information science \cite{Shannon_23} and in physics 
\cite{Jaynes_24,Yukalov_25}. A general convenient form of an information 
functional is given by the Kullback-Leibler relative information 
\cite{Kullback_26,Kullback_27}.

In order to weight the prospects according to their utility, let us consider
a family of prospects $\mathcal{L}$. Assume that they can be weighted by means 
of a distribution defined by {\it utility factors} $f(\pi_j)$ that are normalized,
\be
\label{8}
\sum_{j=1}^L f(\pi_j) = 1 \; , \qquad 0 \leq f(\pi_j) \leq 1 \;   .
\ee
By definition, the utility factor of zero utility is to be zero,
\be
\label{9}
 f(\pi_j) = 0 \; , \qquad U(\pi_j) = 0 \;  .
\ee
Since the utility factors $f(\pi_j)$ weight the finite utilities $U(\pi_j)$, 
the total finite expected utility defined by
\be
\label{10}
 U = \sum_{j=1}^L U(\pi_j) f(\pi_j) \;  
\ee
should exist, given a finite number $L$ of prospects.

Under these conditions, we can define the Kullback-Leibler information as
\be 
\label{11}
 I[f] = \sum_j f(\pi_j) \ln\; \frac{f(\pi_j)}{f_0(\pi_j)} \; + \;
\lbd \left [ \sum_j f(\pi_j) -1 \right ] \; - \;
\bt \left [ \sum_j U(\pi_j) f(\pi_j) - U \right ] \;  ,
\ee
with a trial distribution $f_0(\pi_j)$ proportional to the expected utility 
$U(\pi_j)$ in order to take into account condition (\ref{9}). The parameters 
$\lambda$ and $\beta$ are the Lagrange multipliers guaranteeing the validity 
of the imposed constraints (normalization (\ref{8}) and existence of a well-defined
finite expected utility (\ref{10})) .

Minimizing the information functional (\ref{11}) yields the utility factor
\be
\label{12}
f(\pi_j) = \frac{U(\pi_j)}{Z} \; \exp\{ \bt U(\pi_j) \} \;   ,
\ee
with a normalization coefficient
$$
 Z = \sum_j U(\pi_j) \exp \{ \bt U(\pi_j) \} \;  .
$$
The parameter $\beta$ characterizes the level of belief or confidence of the
decision maker in the correct selection of the prospect set. Requiring that the 
utility factor, by its definition, be an increasing function of utility makes
the belief parameter non-negative $(\beta \geq 0)$.  

In the case of no confidence in the given set of prospects, we have
\be
\label{13}
 f(\pi_j)  = \frac{U(\pi_j)}{\sum_j U(\pi_j) } \qquad (\bt = 0 ) \; .
\ee
In the opposite case of absolute confidence, we get 
\begin{eqnarray}
\label{14}
f(\pi_j) = \left \{ \begin{array}{ll}
1 , & ~~ \pi_j = \max_j \pi_j \\
0 , & ~~ \pi_j < \max_j \pi_j 
\end{array}  \right.
\qquad (\bt \ra \infty) \; ,
\end{eqnarray}
where $\max_j \pi_j$ is the prospect whose expected utility
is the largest. Thus, the latter situation $(\bt \ra \infty )$ retrieves the 
deterministic utility theory, which hence can be seen as a particular case of the 
more general probabilistic approach.  

The prospect utility factors $\{f(\pi_j): j=1, 2, ..., L\}$ give the fractions 
of decision makers selecting the corresponding prospects $\{\pi_j: j=1, 2, ..., L\}$. 
The ordering of prospects in the probabilistic approach is the same as in the 
standard expected utility theory. But now, not all subjects are forced to choose the 
most useful prospect, though it is the prospect whose choice is the most probable.
There can exist a fraction of decision makers choosing other prospects with lower 
utility. The probabilistic decision making scheme is summarized in Fig. 2.

\section{Human decision making and computer operations}

It is widely believed that the human brain operates, during a decision making process,
as a complex and powerful computer. The network of neurons within the brain accepts 
external signals and transforms them into decisions of the subject by accomplishing 
the corresponding actions \cite{Mainzer_28}. Such a procedure could correspond to the 
schemes depicted in Fig. 1 or Fig. 2.  

However, if the brain would act as just described, this would correspond to making 
decisions only on the basis of a well defined deterministic objective function, called 
utility. But there exist numerous empirical studies demonstrating that humans often 
deviate from and even contradict the choices prescribed by utility theory. Such 
contradictions are known as decision-making paradoxes. As examples, we can mention the 
Allais paradox \cite{Allais_29}, the Ellsberg paradox \cite{Ellsberg_30}, the 
Kahneman-Tversky paradox \cite{Kahneman_31}, the Rabin paradox \cite{Rabin_32}, the 
Ariely paradox \cite{Ariely_33}, the disjunction effect \cite{Tversky_34}, the 
conjunction fallacy \cite{Tversky_35}, the planning paradox \cite{Kydland_36}, and many 
others \cite{Camerer_37,Machina_38}. These paradoxes cannot be resolved by the approaches 
consisting in modifying the expected utility theory into so-called non-expected utility 
theories, as has been proved in Refs. \cite{Safra_39,Alnajjar_40}. 

The appearance of numerous paradoxes in decision making, based on utility theory, 
is caused by the fact that this theory does not take into account the emotional
components always present in decision makers, which often compete and modify the decisions 
that would result purely from utility-based processes. A human decision maker not merely 
evaluates the objective utility of the prospects, but also is influenced by subjective 
feelings, emotions, and behavioral biases that are produced by subconscious brain activity. 
The brain takes decisions by combining (i) the objective knowledge of the prospect utility, 
by evaluating the utility factors, with (ii) the subjective attractiveness of the prospects, 
which is hinted by subconscious feelings. The latter means that, in addition to the utility 
factors measured by conscious logical operations, there should exist attraction factors 
produced by subconscious feelings. Then, the resulting weights of the prospects $p(\pi_j)$ 
are defined not merely by the utility factors $f(\pi_j)$, but are also dependent 
on some attraction factors $q(\pi_j)$. We thus suggest that the correct representation
of the brain function during a decision process is given by the scheme
represented in Fig. 3, which should correct and replace those of Figs. 1 or 2. 

Our theory views the human brain not just a powerful computer accomplishing 
a great number of straightforward logical operations, but as an object that 
must include parallel functioning on two levels. One part, representing 
conscious logical operations evaluating the utility factors, can be organized 
as a powerful computer. And the other part, representing subconscious activity 
producing the attraction factors, should be a very different device that functions 
not as a straightforward computer calculating numbers, but as an object 
estimating qualitative features of the prospects.     
       
In the sequel, we do not touch on the technical issues of how the devices, discussed 
above, are actually structurally realized, or how they could be constructed
in an artificial brain. Instead, we describe how their functioning can be represented 
mathematically, characterizing the split dual action of evaluating the prospect 
utilities and estimating their attractiveness.

\section{Quantum decision making by human brain}

The dualism of the brain, combining objective conscious operations with 
subjective subconscious activity, suggests that its functions could be
described by generalizing the real-valued way of defining the prospect weights
to an approach involving complex-valued quantities. In turn, this immediately
points to quantum-theory techniques, where the probability weights are defined
through complex-valued quantities, such as wave functions. 

The idea of employing quantum theory for describing brain functions was
advanced by Bohr \cite{Bohr_41,Bohr_42}. Analyzing the quantum theory of 
measurements, von Neumann \cite{Neumann_43} mentioned that the action of measuring 
observables could be interpreted, to some extent, as taking decisions. Using these 
ideas, we have developed 
\cite{Yukalov_44,Yukalov_45,Yukalov_46,Yukalov_47,Yukalov_48,Yukalov_49}
the Quantum Decision Theory (QDT), using the mathematical techniques of quantum 
theory of measurements. 

Before formulating this theory, we would like to stress that the quantum 
approach to describing human decision making does not assume that the brain is
a quantum object. The quantum techniques just provide the most straightforward 
way of generalizing decision making by taking into account the dual functioning 
of the human brain.       

The main points of QDT are as follows. We consider a set of elementary prospects,
represented by vectors $|n \rgl$, whose closed linear envelope
\be
\label{15}
\cH \equiv {\rm Span} \{ | n \rgl \}
\ee
composes the space of mind. The prospects $\pi_j$ from the given set 
$\mathcal{L}$ are represented by the vectors $|\pi_j \rgl$ in the space of mind.
The prospect operators are defined as
\be
\label{16}
 \hat P(\pi_j) \equiv | \pi_j \rgl \lgl \pi_j | \;  .
\ee
These operators play the same role as the operators from the algebra of local 
observables in quantum theory.

The state of a decision maker is characterized by a non-negative operator $\rho$
acting on the space of mind and normalized as
$$
{\rm Tr} \hat\rho = 1 \;   ,
$$ 
with the trace taken over the space of mind. Defining the decision-maker state by 
a statistical operator, but not by a simple wave function, takes into account that 
this decision maker is not an absolutely isolated subject, but can be influenced by 
its environment.    
  
The prospect probabilities, playing the role of observable quantities, are defined 
as the averages of the prospect operators
\be
\label{17}
 p(\pi_j) \equiv {\rm Tr} \hat\rho \hat P(\pi_j) \;  ,
\ee
with the trace again taken over the space of mind. Writing down the explicit expression 
for the trace over the elementary prospect states and separating the diagonal and 
off-diagonal parts leads to the sum
\be
\label{18}
 p(\pi_j) = f(\pi_j) + q(\pi_j) \;  ,
\ee
in which the first term comes from the diagonal part and the second term, from 
the off-diagonal part. The first term represents the classical utility factor, 
while the second term, caused by the prospect quantum interference, is the 
attraction factor. By definition, the prospect probability is non-negative and 
normalized, so that 
\be
\label{19}
  \sum_{j=1}^L p(\pi_j) = 1 \;  \qquad 0 \leq p(\pi_j) \leq 1 \; .
\ee
In view of the normalization condition for the utility factor (\ref{8}), the 
attraction factor lies in the range
\be
\label{20}
-1 \leq q(\pi_j) \leq 1
\ee
and satisfies the {\it alternation law}
\be
\label{21}
  \sum_{j=1}^L q(\pi_j) = 0 \;  .
\ee

Generally, the attraction factor is a contextual quantity that can vary for
different decision makers and even for the same decision maker at different 
times. This looks as an obstacle for the ability to give quantitative 
predictions for the prospect probabilities. However, it is possible to show
\cite{Yukalov_46,Yukalov_49} that the aggregate attraction factor, averaged
over many decision makers, enjoys the property called {\it quarter law}:
\be
\label{22}
 \frac{1}{L} \; \sum_{j=1}^L | q(\pi_j) | = \frac{1}{4} \;  .
\ee
Since the utility factor is uniquely defined by the corresponding expected 
utility, it is possible to estimate {\it quantitatively} the prospect 
probabilities, assuming that the typical attraction factor satisfies the 
quarter law.      

When the decision maker is a member of a society from which he/she gets 
additional information, then the attraction factor varies depending on the 
amount $\mu$ of the received information. The attraction factor, as a 
function of the information measure $\mu$, can be presented \cite{Yukalov_50} 
in the form 
\be
\label{23}
 q(\pi_j,\mu) = q(\pi_j,0) e^{-\gm\mu} \;  .
\ee
The information can be positive, with $\mu > 0$ as well as negative, or 
misleading, with $\mu < 0$. Respectively, the attraction factor can either 
decrease or increase. The attenuation of behavioral biases with the receipt 
of additional information has been confirmed by empirical studies 
\cite{Kuhberger_51,Charness_52}. 

The reduction of QDT to the probabilistic variant of classical decision theory
corresponds to the attraction factor tending to zero. This is similar to the
reduction of quantum theory to classical statistical theory in the process 
of decoherence.

\section{Cooperation paradox in prisoner dilemma games} 

Let us briefly summarize the status of QDT with respect to its empirical support.
First, the disjunction effect, studied in different forms in a variety of 
experiments \cite{Tversky_34}, has been analyzed in details in 
\cite{Yukalov_46,Yukalov_49}, where we found that the empirically determined 
absolute value of the aggregate attraction factor $|q(\pi_j)|$ was found to coincide 
with the value $0.25$ predicted by expression (\ref{22}), within the typical 
statistical error of the order of $20\%$ characterizing these experiments.
The same agreement, between the QDT prediction for the absolute value of the attraction 
factors and empirical values, holds for experiments testing the conjunction fallacy.
The planning paradox has also found a natural explanation within QDT
\cite{Yukalov_45}. Moreover, it has been shown \cite{Yukalov_48} that QDT explains 
practically all typical paradoxes of classical decision making, arising when 
decisions are taken by separate individuals.

In order to illustrate how QDT resolves classical paradoxes, let us consider
a typical paradox happening in decision making. In game theory, there is a 
series of games, in which several subjects can choose either to cooperate with 
each other or to defect. Such setups have the general name of prisoner 
dilemma games. The cooperation paradox consists in the real behavior of game 
participants who often incline to cooperate despite the prescription of utility 
theory for defection. Below, we show that this paradox is easily resolved within QDT,
which gives correct {\it quantitative predictions}.  

The generic structure of the prisoner dilemma game is as follows. Two 
participants can either cooperate with each other or defect from cooperation. 
Let the cooperation action of one of them be denoted by $C_1$ and the 
defection by $D_1$. Similarly, the cooperation of the second subject is 
denoted by $C_2$ and the defection by $D_2$. Depending on their actions, the 
participants receive payoffs from the set
\be
\label{24}
\mathbb{X} = \{ x_1,x_2,x_3,x_4 \} \;   ,
\ee
whose values are arranged according to the inequality
\be
\label{25}
 x_3 > x_1 > x_4 > x_2 \;  .
\ee
There are four admissible cases: both participants cooperate $(C_1 C_2)$, 
one cooperates and another defects $(C_1 D_2)$, the first defects but the 
second cooperates $(D_1 C_2)$, and both defect $(D_1 D_2)$. The payoffs to each 
of them, depending on their actions, are given according to the rule
\begin{eqnarray}
\left [
\begin{array}{cc}
C_1C_2 & C_1D_2 \\
D_1C_2 & D_1D_2 \end{array} \right ] \ra
\left [
\begin{array}{cc}
x_1x_1 & x_2x_3 \\
x_3x_2 & x_4x_4 \end{array} \right ] \; .
\end{eqnarray}
As is clear, the enumeration of the participants is arbitrary, so that it 
is possible to analyze the actions of any of them. 

Each subject has to decide what to do, to cooperate or to defect, when he/she
is not aware about the choice of the opponent. Then, for each of the 
participants, there are two prospects, either to cooperate,
\be
\label{27}
 \pi_1 = C_1 (C_2 + D_2) \;   ,
\ee
or to defect,
\be
\label{28}
 \pi_2 = D_1 (C_2 + D_2) \;  .
\ee
In the absence of any information on the action chosen by the opponent, 
the probability for each of these actions is $1/2$ (non-informative prior). 
Assuming for simplicity the linear utility as a utility function of the payoffs, 
the expected utility of cooperation for the first subject is
\be
\label{29}
 U(\pi_1) = \frac{1}{2} \; x_1 + \frac{1}{2} \; x_2 \;  ,
\ee
while the utility of defection is 
\be
\label{30}
U(\pi_2) = \frac{1}{2} \; x_3 + \frac{1}{2} \; x_4 \;    .
\ee
The assumption of linear utility is not crucial, and can be removed
by reinterpreting the payoff set (\ref{24}) as the utility set.
Because of condition (\ref{25}), the utility of defection is always larger 
than that of cooperation, $U(\pi_2) > U(\pi_1)$. According to utility theory, 
this means that all subjects have always to prefer defection. 

However, numerous empirical studies demonstrate that an essential fraction of 
participants choose to cooperate despite the prescription of utility theory. 
This contradiction between reality and the theoretical prescription constitutes 
the cooperation paradox \cite{Camerer_53,Tversky_54}.   

Considering the same game within the framework of QDT, we have the probabilities 
of the two prospects,
\be
\label{31}
p(\pi_1) = f(\pi_1) + q(\pi_1) \; , \qquad
p(\pi_2) = f(\pi_2) + q(\pi_2) \;   .
\ee
The propensity to cooperation and the presumption of innocence propose that the 
attraction factor for cooperative prospect is larger than that for the defecting 
prospect, that is, $q(\pi_1) > q(\pi_2)$. In view of the alternation law (\ref{21}) 
and quarter law (\ref{22}), we have
\be
\label{32}
 q(\pi_1) = - q(\pi_2) = \frac{1}{4} \;  .
\ee
Hence, we can estimate the considered prospects by the equations
\be
\label{33}
p(\pi_1) = f(\pi_1) + 0.25 \; , \qquad
 p(\pi_2) = f(\pi_2) - 0.25 \;  .
\ee
From here, we see that, even if defection seems to be more useful than cooperation, 
so that $f(\pi_2) > f(\pi_1)$, the cooperative prospect can be preferred
by some of the participants.   

To illustrate numerically how this paradox is resolved, let us take the data from
the experimental realization of the prisoner dilemma game by Tversky and Shafir 
\cite{Tversky_34}. Subjects played a series of prisoner dilemma games, without 
feedback. Three types of setups were used: (i) when the subjects knew that the 
opponent had defected, (ii) when they knew that the opponent had cooperated, and 
(iii) when subjects did not know whether their opponent had cooperated or defected. 
The rate of cooperation was $3\%$ when subjects knew that the opponent had defected, 
and $16\%$ when they knew that the opponent had cooperated. However, when subjects 
did not know whether their opponent had cooperated or defected, the rate of 
cooperation was $37\%$.    

Treating the utility factors as classical probabilities, we have
$$
f(\pi_1) = \frac{1}{2}\; f(C_1|C_2) + \frac{1}{2}\; f(C_1|D_2) \; ,
$$
$$
f(\pi_2) = \frac{1}{2}\; f(D_1|C_2) + \frac{1}{2}\; f(D_1|D_2) \;   .
$$
According to the Tversky-Shafir data,
$$
f(C_1|C_2) = 0.16 \; , \qquad f(C_1|D_2) = 0.03 \;  .
$$
Hence, 
\be
\label{34}
 f(\pi_1) = 0.095 \; , \qquad f(\pi_2) = 0.905 \;  .
\ee
Then, for the prospect probabilities (\ref{33}), we get
\be
\label{35}
p(\pi_1) = 0.345 \; , \qquad p(\pi_2) = 0.655 \;   .
\ee

In this way, the fraction of subjects choosing cooperation is predicted to be 
about $35 \%$. This is in remarkable agreement with the empirical data of $37 \%$ 
by Tversky and Shafir. Actually, within the statistical accuracy of the experiment, 
the predicted and empirical numbers are indistinguishable.   

If we would follow the classical approach, the fraction of cooperators should 
be not larger than $10 \%$ ($f(\pi_1)$), which is much smaller than the observed $37 \%$. 
But in QDT, there are no paradoxes and its predictions are in quantitative 
agreement with empirical observations.

\section{Conclusion}

We have presented the Quantum Decision Theory that we have developed in the
last four years, which is based on combining utility-like calculations with emotional 
influences in the representation of the decision making processes. We have emphasized 
that decision making by humans is principally different from the direct calculations by, 
even the most powerful, computers. This basic difference is in the duality of the human 
decision-making procedure. The brain makes decisions by a parallel processing 
of two different jobs: by consciously estimating the utility of the available 
prospects and by subconsciously evaluating their attractiveness.    

We have shown how the duality of the brain functioning can be adequately represented 
by the techniques of quantum theory. The process of decision making has been described 
as mathematically similar to the procedure of quantum measurement. The self-consistent 
mathematical theory of human decision making that we have been developed contains no 
paradoxes typical of classical decision making. It is important to stress that this 
theory is the first theory allowing for {it quantitative} predictions taking into account 
behavioral biases. 

We stress that the description of the functioning of the human brain by means of 
quantum techniques does not require that the brain be a quantum object, but this approach 
serves as an appropriate mathematical tool for characterizing the conscious-subconscious 
duality of the brain processes. This duality must be taken into account when
one attempts to create an artificial intelligence imitating the human brain. Such 
an artificial intelligence has to be quantum in the sense explained above
\cite{Yukalov_55}.

\vskip 5mm
{\bf Acknowledgment}

\vskip 3mm
The authors are grateful for many discussions with and advice from M. Favre and 
E.P. Yukalova. Financial support form the Swiss National Science Foundation is 
appreciated.

\newpage

\begin{center}
{\Large{\bf Figure captions}}
\end{center}

\vskip 1cm

{\bf Fig. 1}. Scheme of the deterministic decision making process based on the choice 
of the most useful prospect with the largest expected utility. 

\vskip 1cm
{\bf Fig. 2}. Scheme of the probabilistic decision making process based on the evaluation 
of the prospect utility factors characterizing the fraction of decision makers 
choosing the related prospect.

\vskip 1cm
{\bf Fig. 3}. Scheme of human decision making, which is at the basis of the
Quantum Decision Theory proposed here.

\newpage

\begin{figure}[ht]
\centerline{\includegraphics[width=7cm]{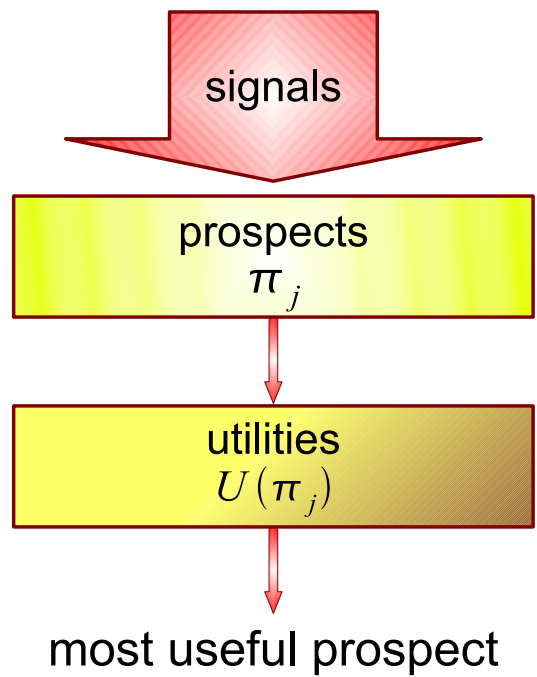} }
\caption{Scheme of the deterministic decision making process based on the choice 
of the most useful prospect with the largest expected utility. }
\label{fig:Fig.1}
\end{figure}

\begin{figure}[ht]
\centerline{\includegraphics[width=10cm]{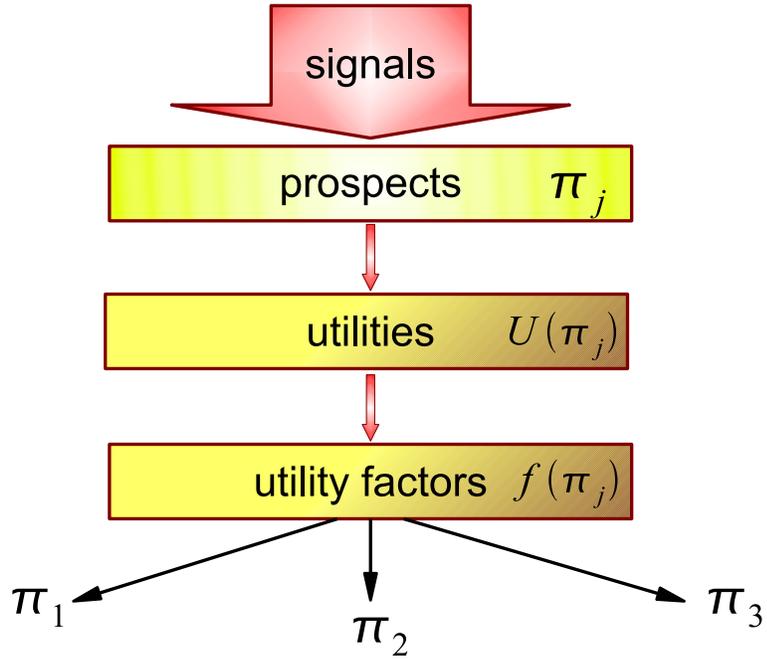} }
\caption{Scheme of the probabilistic decision making process based on the evaluation 
of the prospect utility factors characterizing the fraction of decision makers 
choosing the related prospect.}
\label{fig:Fig.2}
\end{figure}

\begin{figure}[ht]
\centerline{\includegraphics[width=10cm]{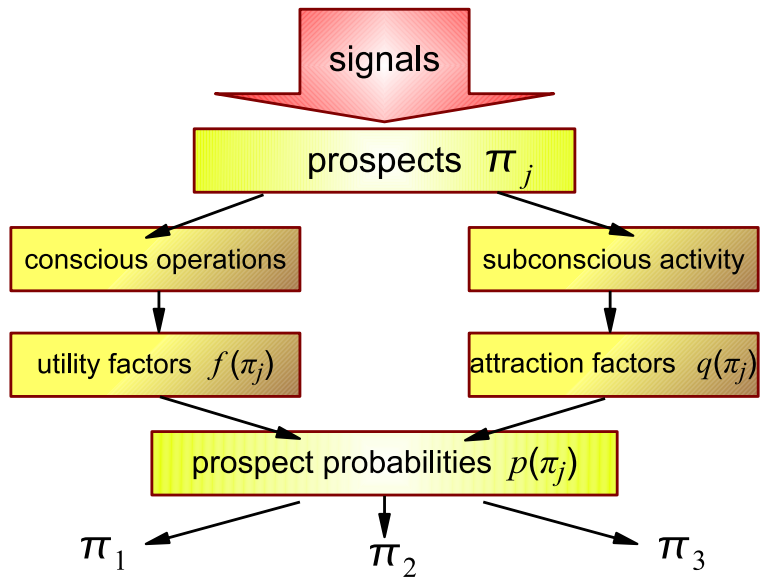} }
\caption{Scheme of human decision making, which is at the basis of the
Quantum Decision Theory proposed here.}
\label{fig:Fig.3}
\end{figure}

\end{document}